\documentclass[a4paper,11pt]{article}
\usepackage{fullpage}
\usepackage{algorithm}

\usepackage{newfloat}
\usepackage{listings}
\usepackage{graphicx}
\usepackage{float}
\usepackage{multirow}

\usepackage[T1]{fontenc}
\usepackage{amsmath,amsthm,amssymb}
\usepackage{tikzsymbols}
\usepackage{multirow}
\usepackage{xcolor}
\usepackage{booktabs}
\usepackage{listings}
\usepackage{subfiles}
\usepackage{csquotes}
\usepackage{makeidx}
\usepackage{tabto}
\usepackage{makecell}
\usepackage{microtype}
\usepackage{enumitem}
\usepackage{todonotes}
\usepackage{pifont}
\usepackage{tikz}
\usetikzlibrary{positioning}
\usepackage{subcaption}
\usepackage{cancel}
\usepackage{tcolorbox}
\usepackage{xfrac}
\usepackage{algorithm}
\usepackage[noend]{algpseudocode}
\usepackage{wrapfig}
\usetikzlibrary{arrows.meta}
\usetikzlibrary{patterns}

\usepackage{pgfplots}

\usepackage{import}

\everymath=\expandafter{\the\everymath\displaystyle}
\usepackage{wrapfig}

\newcommand{\tot}{1131}
\newcommand{\solvdttc}{1112}
\newcommand{\solvdbase}{145}
\newcommand{\solvdvinci}{142}
\newcommand{\accperc}{059}

\newcommand{\R}{\mathbb{R}}

\newcommand{\bR}{\mathbb{R}}
\newcommand{\bN}{\mathbb{N}}
\newcommand{\bZ}{\mathbb{Z}}
\newcommand{\bL}{\mathbb{L}}
\newcommand{\cX}{\mathcal{X}}
\newcommand{\maxinscribed}[1]{\ensuremath{\mathsf{MaxInscribedBall}(#1)}}
\newcommand{\maxratio}[0]{\ensuremath{\mathsf{maxRatio}}}
\newcommand{\getprec}[0]{\ensuremath{\mathsf{GetPrecision}}}
\newcommand{\preci}[0]{\ensuremath{b}}

\newcommand{\vol}[1]{\ensuremath{\mathsf{Volume}\left(#1\right)}}

\newcommand{\ttc}{\ensuremath{\mathsf{ttc}}}
\newcommand{\ttccore}{\ensuremath{\mathsf{ttc}}}

\newcommand{\vinci}{\ensuremath{\mathsf{vinci}}}
\newcommand{\polyvest}{\ensuremath{\mathsf{polyvest}}}

\newcommand{\unif}{\ensuremath{\mathsf{unif}}}
\newcommand{\E}{\ensuremath{\mathbb{E}}}

\newcommand{\numVars}{\ensuremath{n}}
\newcommand{\numCubes}{\ensuremath{m}}
\newcommand{\cube}[1]{c_{#1}}

\newcommand{\eps}{\varepsilon}

\newcommand{\Threshold}{\mathsf{Thresh}}

\newcommand{\pois}{\mathsf{Poisson}}
\newcommand{\GenerateSamples}{\ensuremath{\mathsf{GenerateSamples}}}

\newcommand{\voled}{\ensuremath{\mathsf{ComputeVolume}}}
\newcommand{\polytope}[1]{\ensuremath{\mathsf{Polytope}(#1)}}
\newcommand{\facet}{\textnormal{facet}}

\tikzset{%
  zeroarrow/.style = {-stealth,dashed},
  onearrow/.style = {-stealth,solid},
  c/.style = {circle,draw,solid,minimum width=2em,
      minimum height=2em},
  r/.style = {rectangle,draw,solid,minimum width=2em,
      minimum height=2em}
}

\definecolor{aurometalsaurus}{rgb}{0.43, 0.5, 0.5}
\definecolor{byzantium}{rgb}{0.44, 0.16, 0.39}

\newcommand{\camera}[1]{#1}

\newcommand{\cvc}{\textsc{cvc5}}

\newcommand{\est}{\ensuremath{\mathsf{Est}}}

\newcommand{\sol}[1]{\ensuremath{\mathsf{Sol}(#1)}}

\newcommand{\shsm}[0]{\ensuremath{\mathsf{SharpSMT}}}

\newlist{todolist}{itemize}{2}
\setlist[todolist]{label=$\square$}

\definecolor{Brown}{cmyk}{0,0.81,1,0.60}
\definecolor{OliveGreen}{cmyk}{0.64,0,0.95,0.40}
\definecolor{CadetBlue}{cmyk}{0.62,0.57,0.23,0}
\definecolor{lightlightgray}{gray}{0.9}

\newcommand{\mparagraph}[1]{\par \vspace{.1cm} \noindent \textit{#1}}

\newcommand{\tool}[0]{\ensuremath{\mathsf{ttc}}}

\newcommand{\base}{\polyvest}

\everymath{\textstyle}

\newcommand{\killthis}[1]{}
\newcommand{\Z}{\ensuremath{\mathbb{Z}}}

\newcommand{\OO}[1]{\ensuremath{\mathcal{O}{(#1)}}}

\newcommand{\ceil}[1]{\ensuremath{\lceil #1 \rceil}}

\newcommand{\fl}{\ensuremath{F}}
\newcommand{\fbc}{\ensuremath{{F_{B}^{A}}}}
\newcommand{\mla}{\ensuremath{M}}
\newcommand{\fbd}{\ensuremath{C}}

\newcommand{\todnf}{\ensuremath{\mathsf{toDNF}}}

\newcommand{\ba}{\ensuremath{\mathsf{BooleanAbstraction}}}

\newcommand{\sz}{\widetilde{sz}}
\newcommand{\apxp}{\widetilde{p}}

\newcommand{\hall}{\ensuremath{\mathtt{HALL}}}

\newcommand{\nois}[1]{\ensuremath\mathsf{E}_{#1}^\mathsf{noise}}
\newcommand{\err}[1]{\ensuremath\mathsf{E}_{#1}^\mathsf{error}}
\newcommand{\rnd}[1]{\ensuremath\mathsf{X}_{#1}}

\graphicspath{{./figures/}}

\usepackage{soul}
\usepackage{url}
\usepackage{graphicx}
\usepackage{amsmath}
\usepackage{amsthm}
\usepackage{booktabs}
\usepackage{thmtools}
\pgfplotsset{compat=1.18}
\newtheorem{theorem}{Theorem}
\newtheorem{lemma}{Lemma}

\usepackage{cite}
\usepackage{xcolor}
\usepackage[pdfstartview=FitH,
pdfpagemode=UseNone,
colorlinks,
linkcolor={red!60!black},
citecolor={blue!80!black},
urlcolor={blue!90!black}]{hyperref}
\usepackage[nameinlink]{cleveref}

\newcommand{\citet}[1]{\cite{#1}}
\newcommand{\citeauthor}[1]{\cite{#1}}
\newcommand{\citeyear}[1]{}

\begin{document}
\title{\bf Efficient Volume Computation for SMT Formulas\thanks{  A preliminary version of this work appears at the International Conference on Principles of Knowledge Representation and Reasoning (KR) 2025. The tool {\ttc} is available at \url{https://github.com/meelgroup/ttc}.}}
\author{
\textbf{Arijit Shaw}\\
Chennai Mathematical Institute\\
IAI, TCG-CREST, Kolkata
\and
\textbf{Uddalok Sarkar}\\
Indian Statistical Institute, Kolkata\\
\and
\textbf{Kuldeep S. Meel}\\
Georgia Institute of Technology\\
University of Toronto
}
\date{}

\maketitle

\begin{abstract}
  Satisfiability Modulo Theory (SMT) has recently emerged as a powerful tool for solving various automated reasoning problems across diverse domains. Unlike traditional satisfiability methods confined to Boolean variables, SMT can reason on real-life variables like bitvectors, integers, and reals. A natural extension in this context is to ask quantitative questions. One such query in the SMT theory of Linear Real Arithmetic (LRA) is computing the volume of the entire satisfiable region defined by SMT formulas. This problem is important in solving different quantitative verification queries in software verification, cyber-physical systems, and neural networks, to mention a few.

   We introduce {\ttc}, an efficient algorithm that extends the capabilities of SMT solvers to volume computation. Our method decomposes the solution space of SMT Linear Real Arithmetic formulas into a union of overlapping convex polytopes, then computes their volumes and calculates their union. Our algorithm builds on recent developments in streaming-mode set unions, volume computation algorithms, and AllSAT techniques. Experimental evaluations demonstrate significant performance improvements over existing state-of-the-art approaches.
  \end{abstract}

\section{Introduction}
Satisfiability Modulo Theories (SMT) has revolutionized automated reasoning, serving as the foundational technology for diverse problems~\cite{KS16}. The power of SMT stems from its ability to reason over diverse theories, including bitvectors, reals, and integers, extending well beyond the capabilities of traditional SAT solvers~\cite{cvc5,BSST21,boolector,mathsat5,bitwuzla}. This versatility has established SMT as the de facto decision procedure not only in  formal verification of software and hardware workflows~\cite{HJ20,MMBD+18}, but across numerous domains requiring sophisticated logical reasoning, including security~\cite{BBBB+20}, test-case generation, synthesis, planning~\cite{CMZ20}, and optimization~\cite{SSA16}.

Meanwhile, quantitative reasoning has emerged as a critical advancement in satisfiability solving. Rather than merely determining whether a Boolean formula can be satisfied, model counting~\cite{CMV21,GSS21} techniques calculate the number of satisfying assignments — establishing a robust framework for addressing quantitative challenges like probabilistic inference~\cite{CD08}, software verification~\cite{TW21}, network reliability~\cite{DMPV17}, neural network verification~\cite{BSSM+19}, and numerous other problems~\cite{SM24kr}.

The natural evolution of these parallel developments leads to the compelling extension to effectively handle quantitative queries within SMT frameworks. This challenge is nuanced by the diversity of the underlying theories, each demanding different approaches. In discrete domains like bitvectors and linear integer arithmetic, the problem manifests as model counting. For linear real arithmetic, it transforms into volume computation or counting distinct regions. Recent years have witnessed remarkable progress across these domains, yielding both theoretical insights and practical algorithms for bitvectors~\cite{CMMV16,KM18}, linear integers ~\cite{GB21,G24,GMMZ+19}, reals~\cite{gmpz18,mlz09}, strings~\cite{ABB15}, and projected counting over large classes of SMT formulas~\cite{CDM17,SM25}. Apart from~\cite{gmpz18,mlz09}, these approaches are mostly limited to discrete domains, and hardly extended to continuous domains.  In this work, we address the question: \textit{Given an SMT LRA formula, can we design an efficient volume computation algorithm?}

Our primary contribution is the development of {\ttc}, a novel algorithmic framework that provides an affirmative answer to this question. The {\ttc} algorithm approximates the volume of SMT LRA formula solution spaces with provable theoretical guarantees.

We start with a very related and well-studied problem to SMT volume computation: the problem of volume computation of bounded convex bodies. Sophisticated exact and approximate methods to solve the problem have been developed in the last few decades. Although the exact volume problem is \#P-hard, the seminal work by Dyer, Frieze and Kannan \citet{DFK91} showed a polynomial-time randomized approximation algorithm (FPRAS) for this problem. Furthermore, advancements have not only improved the asymptotic running times of these algorithms~\cite{lovasz1990mixing,AK91,lovasz1991compute,LS92,lovasz1993random,KLS97,LD12,CV18} but also yielded practically efficient methods that forgo certain theoretical guarantees to eliminate prohibitive hidden constants in their runtime~\cite{CV16}.

A central challenge in applying these ideas to SMT lies in the non-convex nature of the solution spaces generated by SMT formulas. Unlike convex bodies, non-convex regions are more complex to analyze due to their irregular shapes and potential discontinuities. A natural strategy to overcome this hurdle is to partition the non-convex space into a union of convex bodies, where each convex piece can be more easily managed with existing techniques.
Decomposing a non-convex SMT solution space into convex components introduces its own set of challenges.  Current state-of-the-art techniques can't handle the union of  non-disjoint components. In many cases, the decomposition yields an excessive number of disjoint components, which may not accurately reflect the underlying structure of the solution space. In practice, solution spaces manifest as unions of overlapping, non-disjoint polytopic regions with boundaries and intersections that encode critical constraint information. This overlapping structure is not merely theoretical—our empirical analysis reveals cases where state-of-the-art decomposition techniques transform a natural representation of 7 overlapping polytopes into an unwieldy collection of 20,595 disjoint components, creating unnecessary computational complexity.

Our approach builds upon recent advancements in counting distinct elements across set unions in streaming models by Meel, Vinodchandran and Chakraborty~\cite{MVC21}.
The fundamental challenge in adapting the MVC algorithm to volume computation arises from the inherent difference between discrete and continuous domains. The MVC algorithm was specifically engineered for discrete settings, while volume computation operates in continuous space. We overcome this obstacle through a principled discretization approach, effectively reducing continuous volume computation to the problem of counting lattice points within a carefully constructed fine-grained lattice space.

We have implemented {\ttc} and evaluated it on a comprehensive benchmark suite. The results demonstrate significant gains in scalability and accuracy. Out of a benchmark set of {\tot} instances, {\ttc} solved {\solvdttc}, while the current state of the art can solve only {\solvdbase}.

\textbf{Applications.}
The theory of linear real arithmetic has significant applications in the formal verification of systems with real variables. These include hybrid systems such as cyber-physical systems~\cite{kdm+23}, and control systems~\cite{CMT12} and timed systems~\cite{mathsat5}. Advanced verification tools like Reluplex~\cite{reluplex} extend SMT solving to neural networks by encoding real-valued variables for network inputs.
Extending these approaches to quantitative verification would require LRA solvers with efficient volume computation capabilities. This follows the established pattern where model counting tools have enabled quantitative verification advances in software~\cite{gfb21,TW21} and binarized neural networks~\cite{BSSM+19}.

\section{Notation and Preliminaries}
\label{sec:background}

An SMT (Satisfiability Modulo Theories) formula $F$ is a quantifier-free logical formula over a background theory $\mathcal{T}$ that may contain both theory atoms (e.g., linear arithmetic predicates) and pure Boolean variables. We focus on formulas over Linear Real Arithmetic (LRA), where theory atoms are of the form $a_1x_1 + \dots + a_nx_n \circ b$, with $a_i, b \in \mathbb{R}$ and $\circ \in \{<, \le, =, \ge, >\}$. 

The Boolean abstraction $F_B$ of $F$ is obtained by replacing each theory atom with a fresh Boolean variable while leaving the original Boolean variables unchanged. We assume that $F_B$ is expressed in Disjunctive Normal Form (DNF) as $F_B = \bigvee_{i=1}^m c_i$, where each cube is given by $c_i = \bigwedge_{j=1}^{n_i} \ell_{ij}$, with each $\ell_{ij}$ being a Boolean literal (either a variable or its negation). And-Inverter Graphs (AIGs) are used as a compact representation for Boolean functions as a circuit, where each node corresponds to a two-input AND gate or an inverter.

We define a mapping $M$ on the Boolean literals corresponding to theory atoms such that for each such literal $\ell$, $M(\ell)$ is its corresponding linear inequality (or its negation). Pure Boolean variables are not mapped, as they do not contribute geometric constraints. For each cube $c_i$, the conjunction $\bigwedge_{j=1}^{n_i} M(\ell_{ij})$ (with the understanding that $M$ is applied only to theory literals) defines a (possibly empty) convex polytope, which we denote by $\mathcal{P}(c_i) = \{ x \in \mathbb{R}^n \mid \forall\, \ell \in c_i \text{ (theory literal)},\, M(\ell)(x) \text{ holds} \}$.

A polyhedron is defined as the intersection of a finite number of half-spaces in $\mathbb{R}^n$, and a polytope is a bounded polyhedron whose volume (measured via the Lebesgue measure) can be computed.
Concretely, any polytope $K$ can be written in the form $\{x \in \mathbb{R}^n \mid A x \leq b\}$, where $A$ is a $m \times n$ matrix and $b$ is a $n \times 1$ vector.
Each row of the system  $A x \;\le\; b$ defines one of the \(m\) half‑spaces (the facets of \(K\)). We denote $\facet(K) = m$, $\sol{K} = \{\,x\in\R^n \mid A x \le b\}$, and $\vol{K}$
as the number of facets, the feasible region, and the volume of \(K\), respectively.

Since we are working over the domain $\bR$, precision is crucial in our work. In our setting, we define precision as the number of digits after the decimal point. 
For example, the number $0.123456789$ has a precision of $9$.

\begin{figure}[!htbp]
  \centering
 {\parbox{\linewidth}{
\begin{lstlisting}[language=c++]
(set-logic QF_LRA)

(declare-const x Real)
(declare-const y Real)

(assert (or
  (and (> x 20) (< x 40) (> y 20) (< y 40))
  (and (> x 10) (< x 30) (> y 10) (< y 30))))

(check-sat)
\end{lstlisting}}}
  \caption{SMT file.}
  \label{fig:ills:formula}
\end{figure}

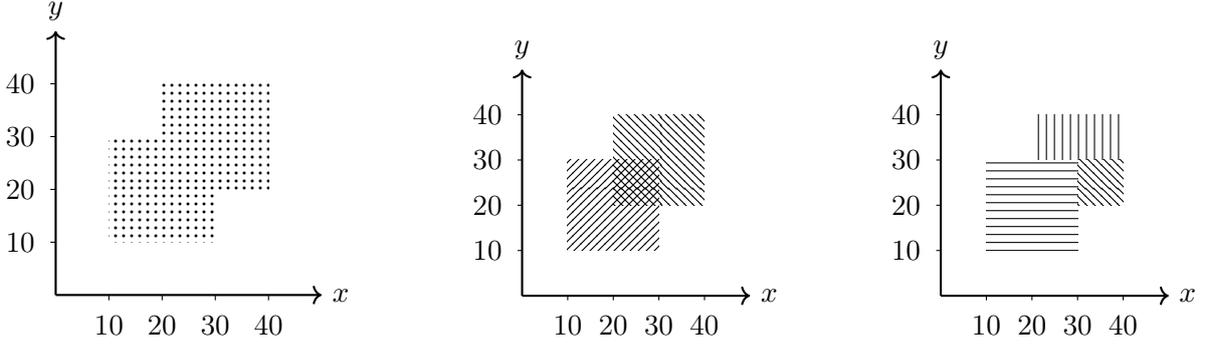
\begin{figure}[!tbp]
    \centering
  \begin{tikzpicture}[x=0.07cm, y=0.07cm]
  \draw[->, thick] (0,0) -- (50,0) node[right] {$x$};
  \foreach \x in {10,20,30,40} {
    \draw (\x,0) -- (\x,-1) node[below=2pt]{\x};
  }

  \draw[->, thick] (0,0) -- (0,50) node[above] {$y$};
  \foreach \y in {10,20,30,40} {
    \draw (0,\y) -- (-1,\y) node[left=2pt]{\y};
  }

  \fill[blue!20, pattern=dots] (10,10) rectangle (30,30);
  \fill[blue!20, pattern=dots] (20,30) rectangle (40,40);
  \fill[blue!20, pattern=dots] (30,20) rectangle (40,30);

\end{tikzpicture}
  \hfill
  \centering
  \begin{tikzpicture}[x=0.06cm, y=0.06cm]
      \draw[->, thick] (0,0) -- (50,0) node[right] {$x$};
      \foreach \x in {10,20,30,40} {
          \draw (\x,0) -- (\x,-1) node[below=2pt]{\x};
 }
      \draw[->, thick] (0,0) -- (0,50) node[above] {$y$};
      \foreach \y in {10,20,30,40} {
          \draw (0,\y) -- (-1,\y) node[left=2pt]{\y};
 }
      \fill[blue!20, pattern=north east lines] (10,10) rectangle (30,30);
      \fill[blue!20, pattern=north west lines] (20,20) rectangle (40,40);
  \end{tikzpicture}
  \hfill
  \begin{tikzpicture}[x=0.06cm, y=0.06cm]
      \draw[->, thick] (0,0) -- (50,0) node[right] {$x$};
      \foreach \x in {10,20,30,40} {
          \draw (\x,0) -- (\x,-1) node[below=2pt]{\x};
 }
      \draw[->, thick] (0,0) -- (0,50) node[above] {$y$};
      \foreach \y in {10,20,30,40} {
          \draw (0,\y) -- (-1,\y) node[left=2pt]{\y};
 }
      \fill[blue!20, pattern=horizontal lines] (10,10) rectangle (30,30);
      \fill[blue!20, pattern=vertical lines] (20,30) rectangle (40,40);
      \fill[blue!20, pattern=north west lines] (30,20) rectangle (40,30);
  \end{tikzpicture}
  \caption{Solution space of SMT formula (left). Disjoint (middle) and non-disjoint (right) decomposition of the solution space of the formula.}
  \label{fig:ills:disndis}
\end{figure}

\begin{figure}[!tbp]
  \centering
  \includegraphics[width=0.4\linewidth]{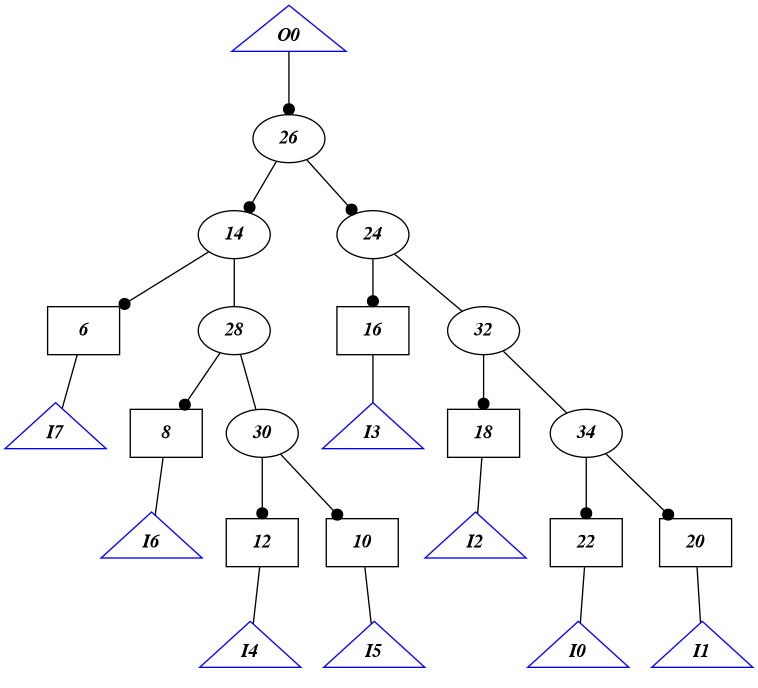}
  \caption{AIG of Boolean abstraction for the example SMT formula, where the variables like $6,8,10,12$ correspond to the LRA atoms $x \leq 10, x \geq 30, y \leq 10, y \geq 30$ respectively.}
  \label{fig:ills:aig}
\end{figure}

\begin{figure}[!tbp]
  \centering
  \resizebox{.4\linewidth}{!}{\input{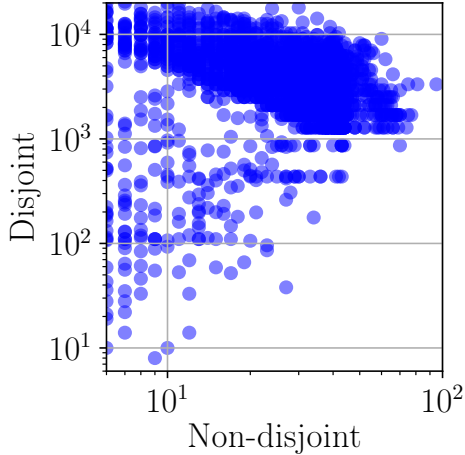}}
  \caption{Comparison of the number of polytopes in disjoint and non-disjoint decomposition (notice the non-equal axis).}
  \label{fig:ills:comparison}
\end{figure}

\mparagraph{Illustrative Example.} Figure~\ref{fig:ills:formula} shows the QF\_LRA formula over two real variables $x$ and $y$, which defines two overlapping square regions in the $xy$-plane. As depicted in Figure~\ref{fig:ills:disndis}, each square has an area of $400$, with an overlapping area of $100$. Thus, the union of the two regions has an area of $700$. Figure~\ref{fig:ills:aig} shows the And-Inverter Graph (AIG) derived from the Boolean abstraction of our linear constraints.
Internal nodes represent logical conjunctions, and any inverted edges (dots) indicate negations. Figure~\ref{fig:ills:disndis}~(left) illustrates a disjoint decomposition of the solution space, where each region is separated so that no two subsets overlap. This often simplifies volume computations but may require more partitions. By contrast, Figure~\ref{fig:ills:disndis}~(right) shows a non-disjoint decomposition, allowing subsets to overlap. Figure \ref{fig:ills:comparison} compares the number of cubes generated by disjoint decomposition and non-disjoint decomposition on our benchmark set. %

\mparagraph{Problem Statement} Let $K$ be the whole solution space of the given formula $F$, which has $d$ dimensions.
Let $\mathcal{B} \subset \mathbb{R}^n$ be an $n$-dimensional measurable set. The volume of $\mathcal{B}$ is defined as $\vol{\mathcal{B}} = \int_{\mathcal{B}} d\mathbf{x}$, where $d\mathbf{x}$ denotes the differential volume element. In Cartesian coordinates, this element is expressed as $d\mathbf{x} = dx_1\,dx_2\,\cdots\,dx_n$.

\section{Related Work}

\mparagraph{SMT Volume Computation} was first addressed by Ma, Liu, and Zhang, who developed the {\vinci} tool~\cite{mlz09} for exact volume computation. Their approach performs intelligent disjoint decomposition of the solution space into convex polytopes using a \textit{bunching} strategy, followed by exact computation of individual polytope volumes. Later, ~\citet{gmpz18} designed {\polyvest}, which extended {\vinci} by incorporating MCMC-based techniques for polytope volume computation. Recently, Ge introduced the {\shsm} tool~\cite{Ge24b}, which integrates and optimizes various techniques from previous work of {\polyvest} and {\vinci}~\cite{gmpz18,mlz09} to create a more comprehensive solution for SMT volume computation.

\mparagraph{Weighted Model Integration} (WMI)~\cite{BPdB15} represents a closely related problem in the hybrid domain of Boolean and rational variables, involving the computation of volume given weight density over the entire domain. This area has seen significant research advances through diverse approaches, including predicate abstraction and All-SMT techniques~\cite{MPS17,MPS19}, knowledge compilation methods~\cite{KMSB+18}, and structurally-aware algorithms~\cite{SMM+22,SMM+24}. However, while WMI focuses primarily on computing weighted integration across domains, our work specifically addresses the fundamental problem of determining the exact volume of the solution space.

\mparagraph{Polytope Volume Computation} has been a central focus in computational geometry since ~\citet{DF88} proved it \#P-hard, followed by Dyer, Frieze, and Kannan's FPRAS development~\cite{DFK91}. Rigorous algorithmic advances have improved complexity bounds from $\widetilde{\mathcal{O}}(n^{23})$ to $\widetilde{\mathcal{O}}(n^3)$~\cite{AK91,KLS97,LV06,LD12,CV18}\footnote{$\widetilde{\mathcal{O}}(\cdot)$ hides the polylogarithmic factors in $\OO{\cdot}$.}. Though these algorithms contain large hidden constants, practical implementations have been achieved by carefully relaxing theoretical guarantees~\cite{CV16,volesti}.

\mparagraph{Estimating the size of a set union} has been a well-studied problem since the work of Karp and Luby~\cite{karp1983monte}, who proposed a $\OO{\numCubes\log^2|\Omega|}$-time algorithm, where $\numCubes$ denotes the number of sets and $|\Omega|$ represents the universe size. Subsequent research has extensively explored streaming settings, where sets arrive sequentially over time~\cite{flajolet1985probabilistic,gibbons2001estimating,kane2010optimal}. More recent methods employ sampling-based strategies in the context of streaming algorithms~\cite{MVC21}, which have also been adapted for DNF counting~\cite{soos2024engineering}. These approaches achieve a $\widetilde{\mathcal{O}}(\numVars\numCubes)$-time complexity for estimating the solution count of a DNF formula, where $\numCubes$ is the number of clauses and $\numVars$ is the number of variables in the DNF formula. Our work builds upon and extends techniques developed in the DNF counting framework.

\camera{
\mparagraph{Volume of a union of polytopes.}
Abboud, Ceylan, and Dimitrov~\cite{ACD20,ACD22} study this problem under a DNF representation of the underlying Boolean formula and compute the volume of the union. Their method builds on the union algorithm of \citet{KLM89}. Our approach is based on MVC, which has shown practical performance advantages over KLM~\cite{soos2024engineering}. Unlike \citet{ACD20}, we permit $\mathcal{O}(\varepsilon)$ error in sampling and volume estimation (vs.\ $\mathcal{O}(\varepsilon^{2}/n)$), where $\varepsilon$ is the volume-approximation factor and $n$ is the dimension.

  }

\section{Algorithm and Analysis}
\label{sec:algo}

This section presents the core contribution of our work: the {\ttc} algorithm along with its theoretical analysis.

\subsection{Algorithm}
Given an SMT LRA formula $\fl$, the {\ttc} algorithm returns an estimate of $\vol{\fl}$.
Initially, the algorithm decomposes the solution space of the SMT formula into non-disjoint polytopes and computes the volume for each polytope. Subsequently, it estimates the volume of the union of the polytopes' solution space using a sampling-based approach.

\begin{algorithm}
    \caption{\ttc $(F,\varepsilon, \delta)$ }
    \label{alg:ttc}
    \begin{algorithmic}[1]
        \State {\fbc},{\mla} $\gets$ \ba({\fl}) \label{line:boolabs}
        \State {\fbd} $\gets$ \todnf({\fbc}) \label{line:allsat}
        \State \preci $\gets$ \getprec({\fbd},$M$,$\frac{\eps}{8}$) \label{line:getprec}
        \State $\eps' \gets \frac{\eps}{12} $, $\delta' \gets \frac{\delta}{2m}$\label{line:eps}
    \State  $\Threshold \gets \max\left(  24  \cdot \frac{\ln (24/\delta)}{(1-\eps')\eps'^2}, 6 (\ln \frac{6}{\delta} + \ln m) \right)$\label{line:thresh}
  \State  $p \gets 1$ ;   $\cX \gets \emptyset$
      \For{$i = 1$ to $\numCubes$}\label{line:final-for-sketch-begin}
          \State $t \gets \voled(\polytope{\cube{i}}, \eps',\delta')$\label{line:volume}
          \For{$s\in \cX$}
              \If{$s \in \polytope{\cube{i}}$}
 remove $s$ from $\cX$\label{line:inside-remove}
              \EndIf
          \EndFor

          \While{$p \geq \frac{\Threshold}{t}$}
              \State Remove every element of $\cX$ with prob. $1/2$ \label{line:prob-remove}
              \State $p \gets p/2$
          \EndWhile
          \State $N_i \gets \pois(t\cdot p)$
          \While{$N_i$ + $|\cX| > \Threshold$}\label{line:final-check-threshold}
              \State Remove every element of $\cX$ with prob. $1/2$ \label{line:final-throw}
              \State $N_i \gets  \pois(t\cdot p/2)$ and $p \gets p/2$ \label{line:final-update-prob}
          \EndWhile
          \State $S \gets \GenerateSamples (\polytope{\cube{i}}, N_i,\preci)$\label{line:sample}
          \State $\cX.\mathsf{Append}$(S)
      \EndFor \label{line:final-for-sketch-end}
      \State Output $|\cX|/p$ \label{line:output}
    \end{algorithmic}
\end{algorithm}

Since we only have a volume computation algorithm for convex polytopes, but the solution space of an SMT formula may be non‐convex, we decompose the solution space as a union of convex polytopes. First, we observe that using $\fbd$, the Boolean abstraction of $\fl$ in DNF form, we can capture the solution space of $\fl$ as a union of polytopes. Let $ {\cube{i}} = \bigwedge_{j=1}^{n_i} \ell_{ij}$ be a cube in the DNF. Then, the conjunction $\bigwedge_{j=1}^{n_i} M(\ell_{ij})$ of the corresponding linear inequalities defines a (possibly empty) convex polytope, which we denote by $\polytope{{\cube{i}}}$. We can show that $\bigcup_{i=1}^{\numCubes} \polytope{{\cube{i}}} = \sol{\fl}$ (\Cref{thm:decomposition}). This relation shows that the DNF representation of the Boolean abstraction of an SMT formula can be used to decompose its solution space into convex polytopes.

To efficiently compute the union of these polytopes, we leverage recent breakthroughs in streaming algorithms for set union operations. The central idea is to compute the union of volumes by maintaining a representative set of points that approximates the total volume. The main caveat of this approach is that the algorithm requires the underlying sets to be finite, while the volume of a polytope cannot be measured directly with a finite number of points. To overcome this issue, we consider an axis-parallel lattice in $\mathbb{R}^n$ with cell side length $10^{-\preci}$, defined as $\bL^n = 10^{-\preci}\mathbb{Z}^n = \{(10^{-\preci}k_1, 10^{-\preci}k_2, \dots, 10^{-\preci}k_n) \mid k_i \in \mathbb{Z} \text{ for } i=1,\dots,n\}$. If $\preci$ is sufficiently large, we can use this lattice to establish a relationship between the number of lattice points contained within a polytope, $|K\cap \bL^n|$, and the volume of the polytope, $\vol{K}$.

We present our algorithm, {\ttc}, in \Cref{alg:ttc}, and in the subsequent part, we describe the algorithm in detail.

In line~\ref{line:boolabs} of \ttc{}, we parse the formula $\fl$ and construct its Boolean abstraction as a circuit, specifically as an And-Inverter Graph (AIG). Then, in line~\ref{line:allsat}, {\ttc} converts the circuit to DNF. The circuit representation enables us to avoid introducing any auxiliary variables. In essence, converting the circuit to DNF is equivalent to solving a circuit AllSAT problem, for which we leverage recent advances in the literature.

Based on the desired accuracy $\eps$ of the volume estimate, we begin by computing the precision parameter $\preci$ using \getprec{} in line~\ref{line:getprec}, and threshold value $\Threshold$ in line~\ref{line:thresh}, which determines approximately how many points will be maintained during the algorithm's execution.
In the main loop (lines~\ref{line:final-for-sketch-begin} to \ref{line:final-for-sketch-end}), we process each cube $\polytope{\cube{i}}$ sequentially. For each cube, we first compute the (approximate) volume of its corresponding polytope using a polytope volume computation algorithm {\voled} (line~\ref{line:volume}). Next, in line~\ref{line:inside-remove}, the algorithm removes from the current set $\cX$ all solutions that are already accounted for. We then determine the number of solutions $N_i$ that would be sampled from $\polytope{\cube{i}}$ if each solution were independently sampled with probability $p$; here, $N_i$ is modeled by a Poisson distribution.
Since we wish to keep the size of $\cX$ bounded by $\Threshold$, if the sum $|\cX| + N_i$ exceeds $\Threshold$, we decrease $p$ and adjust $N_i$ accordingly—this adjustment is performed by resampling $N_i$ from a Poisson distribution with the revised parameter $(p)$ and by removing elements from $\cX$ with probability $1/2$ (line~\ref{line:prob-remove}).
Next, we sample $N_i$ solutions from the cube $\cube{i}$ uniformly at random and add to $\cX$—this is essentially done by uniformly sampling a lattice point from $\polytope{\cube{i}}\cap \bL^n$. Finally, the algorithm outputs the final volume estimate $\frac{|\cX|}{p}$.

\mparagraph{Getting precision.} From ~\citet{KV97}, we know that if an $n$-dimensional $\ell_2$-ball of radius $\gamma = \Omega(n\sqrt{\log f})$ can be inscribed in an $n$-dimensional polytope with $f$ facets, then there exists a relationship between the number of integer lattice points inside the polytope and its volume. When working with the lattice $\bL^n$ instead of the standard integer lattice $\bZ^n$, the effective radius decreases, but the same relationship still applies. Furthermore, as shown in \Cref{lem:nonconv}, the same relationship holds for a union of polytopes, provided each polytope contains an $\ell_2$-ball of radius $\gamma = \Omega\left(n \sqrt{\log \sum_i \facet(P_i)}\right)$.
In \Cref{alg:prec}, we process each polytope sequentially (lines~\ref{line:prec:forstart}--\ref{line:prec:forend}). For each polytope $P^i$, we compute $\hat{\gamma}$, the maximum radius of a ball that can be inscribed in $P^i$, using an LP algorithm (line~\ref{line:prec:getmaxball}). The required precision for this polytope is then determined as $\lceil \log_{10}(\gamma/\hat{\gamma}) \rceil$. After processing all polytopes, we take the maximum of these precision values and return it.

\begin{algorithm}
    \caption{$\getprec$(\fbd,$M$, $\eta$)}
    \label{alg:prec}
    \begin{algorithmic}[1]
        \State \maxratio $\gets 1$, \, $r \gets 0$\label{line:prec:init}
        \For{$\cube{i}$ in {\fbd}}\label{line:prec:forstart}
        \State $P^i \gets \polytope{\cube{i}}$
        \State $r \gets r + \facet(P^i)$
        \EndFor
        \For{$i = 1$ to $m$}\label{line:prec:forstart}
        \State $\hat{\gamma} \gets \maxinscribed{P^i}$ \label{line:prec:getmaxball}
        \State $\gamma \gets \tfrac{16n}{\eta}\sqrt{\log\tfrac{4r}{\eta}}$
        \If{$\frac{\gamma}{\hat{\gamma}} > \maxratio$}
            \State $\maxratio \gets \frac{\gamma}{\hat{\gamma}}$\label{line:prec:forend}
        \EndIf
        \EndFor
        \State \textbf{return} \ceil{\log_{10}(\maxratio)}
    \end{algorithmic}
\end{algorithm}

\subsection{Analysis}

\label{sec:analysis}

\begin{theorem}
  \label{thm:volumettc}
 Given a set $\fl$, and parameters $\eps, \delta \in (0, 1]$, {\ttc} returns an estimate $\est$ of $\vol{F}$ such that with probability at least $1 - \delta$,
  \[
  \est \in \left[(1 - \eps) \cdot \vol{\fl}, \, (1 + \eps) \cdot \vol{\fl} \right].
  \]
\end{theorem}

\begin{theorem}
  \label{thm:decompose}
 The {\ttc} algorithm takes exponential time w.r.t. $v$ in decomposing the polytope, where $v$ is the number of variables in $\fbc$. The number of cubes $m$ can be exponential of $v$ as well.
\end{theorem}
\begin{proof}
 The runtime complexity follows from the dual-rail algorithm of circuit AllSAT algorithm. There exists CNF formulas, for which DNF representation is exponential sized, resulting in the exponential blowup of $m$.
\end{proof}

\begin{theorem}
  \label{thm:runtime}
 Let $ m $ denote the number of decomposed polytopes. Then the {\ttc} algorithm runs in time $\OO{mn^4}$, where $ n $ is the number of dimensions and the $\OO{\cdot}$ notation suppresses polylogarithmic factors in $\eps$ and $\delta$.
\end{theorem}

\begin{proof}
The algorithm's main loop (lines~\ref{line:final-for-sketch-begin}--\ref{line:final-for-sketch-end}) executes exactly $ m $ iterations. In each iteration, the volume computation (line~\ref{line:volume}) requires $\OO{n^4}$ time, while the sampling step (line~\ref{line:sample}) consumes $\OO{n^3}$ time per sample. Since these two operations dominate the computational cost in each iteration, the total runtime is given by $m \times \OO{n^4} = \OO{mn^4}$, where the polylogarithmic dependencies on $\eps$ and $\delta$ are hidden in the $\OO{\cdot}$ notation.
\end{proof}

Now we prove the theorem~\ref{thm:volumettc} using the following lemmas.

\begin{lemma}
  \label{thm:decomposition}
Let $\fbd = \bigvee_{i=1}^{\numCubes} {\cube{i}}$ be the DNF abstraction of the SMT formula $\fl$. Then,
  $$\bigcup_{i=1}^{\numCubes} \polytope{{\cube{i}}} = \sol{\fl}$$
\end{lemma}

\begin{proof}
 First, we show the soundness of each cube: we claim that for every cube ${\cube{i}}$ in $\fbd$, $\polytope{{\cube{i}}} \subseteq \sol{\fl}$. Let $ {\cube{i}} = \bigwedge_{j=1}^{n_i} \ell_{ij} $ be an arbitrary cube of $\fbd$. By construction, we have $\ell_{i,j} \models \fbd$. Therefore, by the property of Boolean abstraction,  if $\bigwedge M(\ell_{i,j})$ can be satisfied, then $\fl$ is satisfied. Any point $x \in \polytope{{\cube{i}}}$ satisfies $\bigwedge M(\ell_{i,j})$,    therefore  $x \in \sol{\fl}$, proving that $ \polytope{{\cube{i}}} \subseteq \sol{\fl}.$

 Next, we show the completeness of the union, that $\sol{\fl} \subseteq \bigcup_{i=1}^{\numCubes} \polytope{{\cube{i}}}$. Consider any arbitrary point $x \in \sol{\fl}$. Now $x$ induces a Boolean assignment satisfying $\fbd$. Since $\fbd = \bigvee_{i=1}^{\numCubes} {\cube{i}}$, there exists at least one cube ${\cube{i}}$ such that $x$ satisfies every literal in ${\cube{i}}$. By the definition of $M$, we have that for every $l \in {\cube{i}}$, $M(l)(x)$ holds, which implies that $x \in \polytope{{\cube{i}}}$. Thus, $\sol{\fl} \subseteq \bigcup_{i=1}^{\numCubes} \polytope{{\cube{i}}}$.
\end{proof}

\begin{lemma}[Lemma 1 of \protect\citet{soos2024engineering}]
  \label{lem:chernoff}
 Let $N$ be random variable such that $N \sim \pois(\lambda)$, for $\lambda > 0$. Then, for any $x > 0$, the following inequalities hold.
  \[\textnormal{(i)} \Pr\left[|N - \lambda| \geq \zeta \right] \leq 2 \exp\left(-{\zeta^2}/{(2\lambda + \zeta)}\right) \] \[ \textnormal{(ii)}\Pr\left[N - \lambda \geq \zeta \right] \leq \exp\left(-{\zeta^2}/{(2\lambda + \zeta)}\right)\]
\end{lemma}

\begin{restatable}{lemma}{process}
  \label{lem:process}
  Let $\beta > \alpha > 0$ and $\eta \in [0,1]$.  
  The following three procedures yield statistically equivalent distributions:
  \begin{enumerate}
      \item Draw $N \sim \pois(\sz \cdot p/2)$ samples with replacement from $S$, where each element is drawn from an $\eta$-close uniform distribution, i.e., $\Pr(s) \in \left[\frac{1-\eta}{|S|},\, \frac{1+\eta}{|S|}\right]$, with $\alpha|S| \leq \sz \leq \beta|S|$ and add them to $\cX$.
      \item Draw $N \sim \pois(\sz \cdot p)$ samples from $S$ using the same $\eta$-close uniform distribution, with $\alpha|S| \leq \sz \leq \beta|S|$, then independently discard each sample with probability $1/2$, adding the rest to $\cX$.
      \item For each $s \in S$, independently draw $r \sim \pois(\apxp/2)$, where $(1-\eta)\alpha p \leq \apxp \leq (1+\eta)\beta p$, and add $r$ copies of the element $s$ to $\cX$.
  \end{enumerate}
\end{restatable}

\begin{lemma}[Theorem 3 of \protect\citet{KV97}]
  \label{lem:integerlattice}
 Let $P$ be a polytope in $\bR^n$ that contains an $\ell_2$-ball of radius at least $8n \sqrt{\log (2\facet(P)/\eta)}$, where $0< \eta \leq 1$ is a constant. Then $\vol{P}$ satisfies the following bounds, \[\vol{P} \in \left[\frac{1-\eta}{2} |P \cap \bZ^n|, \left.\right. (1+\eta) |P \cap \bZ^n|\right]\]
\end{lemma}

We extend this result to the case of the union of polytopes in the following lemma.
\begin{restatable}{lemma}{nonconv}
  \label{lem:nonconv}
 Let \(Q=\bigcup_{i=1}^{m}P_i\) with \(P_i=\polytope{\cube{i}}\).
 If every \(P_i\) contains an \(\ell_{2}\)-ball of radius at least
  \(\tfrac{16n}{\eta}\sqrt{\log \tfrac{4r}{\eta}}\) where \(0<\eta\le1\) a constant and
  \(r = \sum_{i = 1}^m \facet(P_i)\), then
  \[\vol{Q} \in \left[
  (1-\eta)\,|Q\cap\Z^{n}|,
  (1+\eta)\,|Q\cap\Z^{n}|\right]
  \]
\end{restatable}
\begin{proof}
 Recall the description of $P_i$ as
  \(P_i=\{x\in\R^{n}\mid A_{ij}x\le b_{ij}\;(j=1,\dots,\facet(P_i))\}\)
 and set
  \[
    \kappa:=\sqrt{2 \log \frac{4}{\eta}}+\sqrt{2\log r},\quad
 k:=\max_{i,j}\frac{b_{ij}+\kappa||A_{ij}||}{\,b_{ij}-\kappa||A_{ij}||}
  \]
 where \(r=\sum_{i=1}^{m}\facet(P_i)\) and $||\cdot||$ denotes the $\ell_2$-norm.
 Expand/shrink every facet by \(\pm\kappa||A_{ij}||\) and write
  \(P_i':=\{x\mid A_{ij}x\le b_{ij}+\kappa||A_{ij}||\},\;
 P_i'':=\{x\mid A_{ij}x\le b_{ij}-\kappa||A_{ij}||\}\), and let \(Q':=\bigcup_iP_i',\;Q'':=\bigcup_iP_i''\).
 Let $\rnd{}$ be the randomized rounding process as defined in~\citet{KV97}, which takes a point $p \in \bR^n$ and returns a point in the lattice $x \in \Z^n$ such that $|x -p| \leq 1$.
 To complete the proof, we will need the following lemma.

    \begin{restatable}{lemma}{ballclaim}\label{claim:cl}
 The following properties hold for the sets \(Q'\) and \(Q''\):
      \textbf{(a)} Draw $p$ uniformly from the continuous body $Q'$.
 Then for every $x\in Q\cap\Z^{n}$,
      $
        \Pr[\rnd{p}=x]\;\ge\;\frac{1-\eta/2}{\vol{Q'}}.
      $
      \textbf{(b)} Draw $p$ uniformly from the \emph{continuous} body $Q''$.
 The event that $p$ still lies in $Q$ satisfies
      $
        \Pr[\rnd{p}\in Q]\;\ge\;1- \frac{\eta}{2},
      $
 and \textbf{(c)} for every $x\in Q\cap\Z^{n}$,
      $
        \Pr[\rnd{p}=x]\;\le\;\frac{1}{\vol{Q''}}.
      $
    \end{restatable}

    \begin{proof}[Proof]
  To begin with the proof we define random variables $Y$ such that
  We will prove the lemma in two parts. First, we consider that $p$ is drawn uniformly from $Q'$.
  Consider the random variable $Y$ such that $Y \in [-1,1]^n$ and defines the event $\rnd{p} = x$ as $x + Y = p$.
  Then we have,
  \begin{align*}
    \Pr (\rnd{p} = x) &= \int_{p \in Q'}\Pr (\rnd{p} = x \ | \ p ) \cdot \unif(p) dp\\
    &= \int_{p \in Q'}\Pr (\rnd{p} = x \ | \ p ) \cdot \frac{dp}{\vol{Q'}} && \text{Since $\unif$ is uniform over $Q'$}\\
    &= \frac{1}{\vol{Q'}} \int_{p \in Q'}\Pr (x + Y = p) dp && \text{Since $\Pr (x + Y = p)=\Pr (\rnd{p} = x \ | \ p )$}
  \end{align*}
  Now since $\Pr(x + Y =p) = 0$ for $p \notin C(x)$ where $C(x)$ is a cube centered at $x$ with all the sides to be of length $2$. Therefore,
    \begin{align*}
        \int_{p \in Q'}\Pr (x + Y = p) dp &= \int_{p \in C(x)}\Pr (x + Y = p) dp - \int_{p \in C(x)\setminus Q'}\Pr (x + Y = p) dp\\
        &= 1 - \int_{p \in C(x)\setminus Q'}\Pr (x + Y = p) dp
    \end{align*}
    The last equality follows from $\int_{p \in C(x)}\Pr (x + Y = p) dp = 1$, since $Y \in [-1,1]$. Now let us bound $\int_{p \in C(x)\setminus Q'}\Pr (x + Y = p) dp$. Note that, for $x \in Q$, $A_{ij} x \leq b_{ij}$ and $p \in C(x) \setminus Q'$, $A_{ij} p \geq b_{ij} + \kappa ||A_{ij}||$, therefore $A_{ij} Y \geq \kappa||A_{ij}||$. Now for fixed $i,j$, we consider the random variables $Z_k = \sum_{k' = 1}^k A_{ijk'}Y_{k'}$. Since, $\E[Z_k \ | \ Z_{k-1}] = Z_{k-1}$, we have that $Z_k$ is a martingale. Therefore, since $|Z_k - Z_{k-1}| \leq A_{ijk}$, applying the Azuma's inequality, we have:
    $\Pr(A_{ij} Y \geq \kappa) \leq 2 \exp\left({-\frac{\kappa^2||A_{ij}||}{2\sum_{k'=1}^n|A_{ijk'}|^2}}\right) \leq 2 \exp\left(-(\log \frac{4}{\eta} + \log r)\right) = \frac{\eta}{2r}$. Therefore using union bound, we have:
    \begin{align*}
        \int_{p \in C(x)\setminus Q'}\Pr (x + Y = p) dp \leq \sum_{ij} \Pr(A_{ij} Y \geq \kappa) \leq \frac{\eta}{2}
    \end{align*}
    Therefore we have $1-\frac{\eta}{2} \leq \int_{p \in Q'}\Pr (x + Y = p) dp$, which proves the first part (a) of the claim:
    \[\frac{1-\eta/2}{\vol{Q'}} \leq \Pr (\rnd{p} = x)\]
  Similarly, if $p$ is sampled from $Q''$, we have for every $p \in Q''$,
  \begin{align*}
    \Pr (\rnd{p} \notin Q) \leq \sum_{ij} \Pr(A_{ij} Y \geq \kappa) \leq \frac{\eta}{2}
  \end{align*}
  Therefore we have the second part (b) of the claim:
  \[\Pr (\rnd{p} \in Q) \geq \int_{Q''} \frac{1 - \eta/2}{\vol{Q''}} dp \geq 1 - \frac {\eta}{2}\]
  Finally for a fixed $x\in Q\cap\Z^{n}$, we have:
  \allowdisplaybreaks
  \begin{align*}
    \Pr[\rnd{p}=x] &= \int_{p \in Q''}\Pr (\rnd{p} = x \ | \ p ) \cdot \unif(p) dp\\
    &= \int_{p \in Q''}\Pr (\rnd{p} = x \ | \ p ) \cdot \frac{dp}{\vol{Q''}} && \text{Since $\unif$ is uniform over $Q''$}\\
    &= \frac{1}{\vol{Q''}} \int_{p \in Q''}\Pr (x + Y = p) dp && \text{Since $\Pr (x + Y = p)=\Pr (\rnd{p} = x \ | \ p )$}\\
    &\leq \frac{1}{\vol{Q''}} \int_{p \in C(x)}\Pr (x + Y = p) dp && \text{Since in $Q'' \setminus C(x)$ the probability is 0}\\
    &= \frac{1}{\vol{Q''}} && \qedhere
  \end{align*} 
\end{proof}

 We now use this lemma to complete the proof.
 Consider the case where $p$ is drawn uniformly from $Q''$.
 Then using \cref{claim:cl}(b) and \ref{claim:cl}(c) we have:
    \begin{align*}
      1- \frac{\eta}{2} &\leq \sum_{x\in Q\cap\Z^{n}} \Pr[\rnd{p}=x]\\
      &\leq \sum_{x\in Q\cap\Z^{n}} \frac{1}{\vol{Q''}} \, \leq \frac{|Q\cap\Z^{n}|}{\vol{Q}}
    \end{align*}
 The last inequality follows from $Q \subseteq Q''$.
 Since for all $0 < \eta \leq 1$ we have, $\left(1 - \frac{\eta}{2}\right)^{-1} \leq (1+\eta)$, therefore,
    $\vol{Q}\leq (1+\eta)|Q\cap\Z^{n}|$.
 Again, consider the case where $p$ is drawn uniformly from $Q'$. Then, using \cref{claim:cl}(a)
    \begin{align*}
      1 &\geq \sum_{x\in Q\cap\Z^{n}} \Pr[\rnd{p}=x] \\
      &\geq \sum_{x\in Q\cap\Z^{n}} \frac{1-\eta/2}{\vol{Q'}} \, = \left(1-\frac{\eta}{2}\right)\cdot\frac{|Q\cap\Z^{n}|}{\vol{Q'}}
    \end{align*}
 Now consider $p \in Q'$. Then $A_{ij}p \leq b_{ij}'$ for all $i,j$ and therefore $A_{ij} \frac{p}{k} \leq b_{ij}''$ for all $i,j$. This implies that $\frac{p}{k} \in Q''$. This implies that $Q' \subseteq kQ''$. Therefore, $\vol{Q'} \leq k^n \vol{Q''}$. Since each $P_i$ contains an \(\ell_2\)-ball of radius at least $\tfrac{16n}{\eta}\sqrt{\log \tfrac{4r}{\eta}}$, it follows that for the pair $i,j$ at which $k$ is attained, we have $b_{ij} \geq \tfrac{16n}{\eta} \sqrt{\log \tfrac{4r}{\eta}}||A_{ij}||$. Consequently,
 \[k^n = \left(\frac{\tfrac{16n}{\eta} \sqrt{\log \tfrac{4r}{\eta}} + \kappa}{\tfrac{16n}{\eta} \sqrt{\log \tfrac{4r}{\eta}} -  \kappa}\right)^n \]
 \begin{restatable}{claim}{helper}
 \label{claim:helper}
  For $n \in \bN$ we have,
  \[
    k^n \leq \frac{1+\eta /4}{1-\eta/4}
  \]
 \end{restatable}

Using Claim \ref{claim:helper}, we can upper bound $\vol{Q'}$ as follows: $\vol{Q'} \leq \frac{1+\eta/4}{1-\eta/4} \cdot \vol{Q''} \leq \frac{1+\eta/4}{1-\eta/4} \cdot \vol{Q}$. Using the inequality $\frac{1-\eta/4}{1+\eta/4} \cdot \left(1-\frac{\eta}{2}\right) \ge (1-\eta)$ for all $0 < \eta \leq 1$, we have $\vol{Q} \ge (1-\eta)|Q \cap \bZ^n|$, completing the proof of the lemma.
\end{proof}

\begin{lemma}
  \label{lem:lattice}
 Let \(Q=\bigcup_{i=1}^{m}P_i\) with \(P_i=\polytope{\cube{i}}\) such that every $P_i$ follows the condition of \cref{lem:nonconv}, then we have the following bound on the number of lattice points in $Q$, \[\vol{Q} \in \left[\frac{1-\eta}{10^{\preci\cdot n}} |Q\cap \bL^n|, \left. \right . \frac{1+\eta}{10^{\preci\cdot n}}|Q\cap \bL^n|\right]\]
 where $\preci$ is the precision parameter selected by $\getprec$.
\end{lemma}
\begin{proof}
 We define a canonical isomorphism $\phi: \bL^n \to \bZ^n$. For convenience, we use the same notation to denote the image of a polytope $P_i$ (resp. $Q$) under $\phi$, writing $\phi(P_i)$ (resp. $\phi(Q)$).
 This mapping scales the distance between any two consecutive points $[a,b]$ along a dimension by $10^\preci$. Consequently, a ball of radius $r$ in $\bL^n$ corresponds to a ball of radius $10^\preci r$ in $\bZ^n$. Furthermore, the volume of a polytope $P_i$ (resp. $Q$) scales by a factor of $10^{\preci\cdot n}$ when transformed into $\bZ^n$, leading to $\vol{P_i} \cdot 10^{\preci\cdot n}$ (resp. $\vol{Q} \cdot 10^{\preci\cdot n}$). Since the mapping $\phi$ preserves the lattice structure, we have $|\phi(P_i)\cap \bZ^n| = |P_i\cap \bL^n|$ and $|\phi(Q)\cap \bZ^n| = |Q\cap \bL^n|$.

 Finally, the choice of $\preci$ as selected by $\getprec$ guarantees that each $P_i$ contains a ball of radius at least $10^{-\preci}\tfrac{16n}{\eta} \sqrt{\log (4\sum_{i=1}^m\facet(P_i)/\eta)}$. Therefore, we can apply \cref{lem:nonconv} to complete the proof.
\end{proof}

\begin{lemma}
  \label{lem:numlattice}
 Let \( Q = \bigcup_{i=1}^{m} \polytope{\cube{i}} \). Then {\ttccore} outputs an estimate \(\est\) such that with probability at least \(1 - \delta\),
  \[
  \est \in \left[\frac{(1 - \eps/3)}{10^{\preci \cdot n}} \left| Q \cap \bL^n \right|, \,\frac{(1 + \eps/3)}{10^{\preci \cdot n}} \left| Q \cap \bL^n \right|\right]
  \]
\end{lemma}

\begin{proof}
  Let $\eps' = \frac{\eps}{12}$ as defined in \cref{line:eps}. 
  We denote the polytope corresponding to $\cube{i}$ as $P^i = \polytope{\cube{i}}$. Let $S^{(i)}$ represent the set of lattice points within all polytopes up to index $i$, that is, $S^{(i)} = \cup_{j\leq i} (P^j \cap \bL^n)$. We also define, $\cX^{(i)}_j$ as the state of $\cX$ after consuming $P^i$ when the value of $p$ is $\frac{1}{2^j}$.
  Before proceeding with the proof, we introduce the following key invariant:
  \begin{quote}
    \textbf{Invariant} For any $i \in [\numCubes]$ and $j \geq 0$ the multiset $\cX^{(i)}_j$ contains $k$ copies of each lattice point inside $S^{(i)}$, where $k$ is a random variable drawn from $\pois(\apxp_j)$ such that $\frac{(1-\eps')(1-\eps/8)}{2^j10^{\preci \cdot n}} \leq \apxp_j \leq \frac{(1+\eps')(1+\eps/8)}{2^j10^{\preci \cdot n}}$.
  \end{quote}
  \Cref{lem:lattice} ensures that for any $P^i$, $\vol{P^i} \in \left[\frac{(1-\eps/8)}{10^{\preci\cdot n}} |P^i\cap \bL^n|, \left. \right . \frac{(1+\eps / 8)}{10^{\preci\cdot n}}|P^i\cap \bL^n|\right]$, and by the accuracy guarantee of \voled, the call $\voled(P^i, \eps', \delta')$ returns $t$ such that $t \in \left[(1-\eps')\vol{P^i}, (1+\eps')\vol{P^i}\right]$.
  Therefore, the invariant follows from \cref{lem:process}.
  Now let us define two events corresponding to the run of the algorithm:
  \begin{enumerate}
    \item $\nois{j}$: After consuming the last polytope $P^\numCubes$, the value of $p$ is $\frac{1}{2^j}$.
    \item $\err{j}$: At the end of the algorithm $\cX^{(\numCubes)}_j \notin \left[(1-\eps')\apxp_j|Q\cap \bL^n|, (1+\eps')\apxp_j|Q\cap \bL^n|\right]$.
  \end{enumerate}
  Noting that $(1-\eps')\apxp_j \geq \frac{(1-\eps/3)}{2^j10^{\preci \cdot n}}$ and $(1+\eps')\apxp_j \leq \frac{(1+\eps/3)}{2^j10^{\preci \cdot n}}$ for all $0\leq\eps \leq1$, it suffices to establish the following bound is enough to complete the proof of the lemma:
  \[
    \Pr\left[\bigcup_{j=0}^{\infty} (\nois{j} \land \err{j})\right] \leq \delta.
  \]
  Let $j^*$ be the smallest $j$ such that $\frac{1}{2^j 10^{\preci \cdot n}} < \frac{\Threshold}{4 |S^{(\numCubes)}|}$. Applying union bound, we get
  \[\Pr\left[\bigcup_{j=0}^{\infty} (\nois{j} \land \err{j})\right] \leq \sum_{j=0}^{j^*-1}\Pr\left[\err{j}\right] + \Pr\left[\cup_{j\geq j^*}\nois{j}\right] \]
  We now bound these terms separately.

  \textbf{Bounding $\Pr\left[\cup_{j\geq j^*}\nois{j}\right]$}:
  For $j < j^*$, we have $\frac{1}{2^j10^{\preci \cdot n}} \geq \frac{\Threshold}{4|S^{(\numCubes)}|}$. Since the invariant ensures that $\cX^{(\numCubes)}_j $ contains $k$ copies of each lattice point in $S^{(\numCubes)}$, where $k \sim \pois(\apxp_j)$, Therefore, by \cref{lem:process}, we have that $\cX^{(\numCubes)}_j$ follows $\pois(|S^{(\numCubes)}|\apxp_j)$ distribution. The event $\cup_{j \geq j^*}\nois{j}$ occurs if $p$ decreases from $\frac{1}{2^{j^*-1}}$ to $\frac{1}{2^{j^*}}$ for some $i \leq \numCubes$. This is equivalent to the case that $\cX > \Threshold$ for some $\cX \sim \pois(\frac{|S^{(i)}|}{2^{j^*-1}10^{\preci \cdot n}})$. Since from the choice of $j^*$ we have, $\Threshold > \frac{4|S^{(\numCubes)}|}{2^{j^*}10^{\preci \cdot n}}$, therefore $\Threshold - \frac{|S^{(i)}|}{2^{j^*-1}10^{\preci \cdot n}} > \frac{\Threshold}{2}$. Thus, $\Pr[\cX > \Threshold]$ can be bounded as follows:
  \[
    \Pr\left[\cX - \frac{|S^{(i)}|}{2^{j^*-1}10^{\preci \cdot n}} > \Threshold - \frac{|S^{(i)}|}{2^{j^*-1}10^{\preci \cdot n}}\right] \leq \Pr\left[\cX - \frac{|S^{(i)}|}{2^{j^*-1}10^{\preci \cdot n}} > \frac{\Threshold}{2}\right]
  \]
  Applying \cref{lem:chernoff}, for any $i \leq \numCubes$ we have, $\Pr[\cX > \Threshold] \leq \exp\left(-\frac{\Threshold}{6}\right) \leq \frac{\delta}{6\numCubes}$. The last inequality follows from $\Threshold \geq 6\log (6\numCubes/\delta)$.
  Therefore, by union bound over $i\in [\numCubes]$, $\Pr\left[\cup_{j\geq j^*}\nois{j}\right] \leq \frac{\delta}{6}$.

  \textbf{Bounding $\sum_{j=0}^{j^*-1}\Pr\left[\err{j}\right]$}:
  By noting that $\cX^{(\numCubes)}_j$ follows $\pois(|S^{(\numCubes)}|\apxp_j)$ distribution, therefore using \cref{lem:chernoff} we have that for any $j < j^*$, \[\Pr\left[\err{j}\right] \leq 2 \exp\left(-\frac{\eps'^2 |S^{(\numCubes)}| \apxp_j}{3}\right)\]
  Now from the definition of $j^*$ we have $\frac{|S^{(\numCubes)}|}{2^{j^*-1}10^{\preci \cdot n}}> \frac{\Threshold}{4}$, hence $|S^{(\numCubes)}| \apxp_{j^*-1}> \frac{(1-\eps')\Threshold}{8}$. Consequently, \[\Pr\left[\err{j}\right] \leq 2 \exp\left(-\frac{\eps'^2 |S^{(\numCubes)}| \apxp_{j^* -1}}{3}\right)\leq 2 \exp\left(-\frac{(1-\eps')\eps'^2 \Threshold}{24}\right)\]
  Therefore, we have that,
  \begin{align*}
    \sum_{j=0}^{j^*-1}\Pr\left[\err{j}\right] &\leq 2 \exp\left(-\frac{(1-\eps')\eps'^2 \Threshold}{24}\right) + 2 \exp\left(-\frac{(1-\eps')\eps'^2 \Threshold}{12}\right) + \ldots \\
    &\leq 4 \exp\left(-\frac{(1-\eps')\eps'^2 \Threshold}{24}\right) \leq \frac{\delta}{6}
  \end{align*}
  The last inequality follows from $\Threshold \geq \frac{24\log(24/\delta)}{(1-\eps')\eps'^2}$.
\end{proof}

Now we finish the proof of \cref{thm:volumettc} by combining the results from \cref{lem:numlattice} and \cref{lem:lattice}.

\begin{proof}[Proof of \cref{thm:volumettc}]
  Using \cref{lem:numlattice}, we have $\est \geq \frac{1 - \eps/3}{10^{\preci \cdot n}} \cdot |Q \cap \bL^n|$, and from \cref{lem:lattice}, $\frac{|Q \cap \bL^n|}{10^{\preci \cdot n}} \geq \frac{\vol{Q}}{(1 + \eps/8)}$.
  Combining these two inequalities, we get
   $
   \est \geq \frac{1 - \eps/3}{1 + \eps/8} \cdot \vol{Q} \geq (1 - \eps) \cdot \vol{Q}.
   $
  Similarly, the corresponding upper bounds from \cref{lem:numlattice} and \cref{lem:lattice} we yield
   $
   \est \leq \frac{1 + \eps/3}{1 - \eps/8} \cdot \vol{Q} \leq (1 + \eps) \cdot \vol{Q}.
   $
 \end{proof}

\subsection{Implementation}
We implemented a Python prototype for testing the algorithm and its efficiency. We use the following tools for different parts of the \Cref{alg:ttc}.

\mparagraph{Decomposing into Polytopes.} We use a combination of the existing SMT solver and Circuit AllSAT solver to decompose the SMT solution space into convex polytopes. Specifically, we do the following:

\mparagraph{Boolean Abstraction.}
To create the Boolean abstraction of the SMT formula in line \ref{line:boolabs} of \Cref{alg:ttc}, we instrument {\cvc}~\cite{cvc5} to create the abstraction as an AIG. By default, {\cvc} creates the abstraction as a CNF, which we wanted to avoid, since converting to CNF introduces numerous auxiliary variables, making it difficult to enumerate the solutions in a DNF form. For this purpose, we carefully examined each different type of Boolean operation used in SMT formulas, which is typically translated to CNF, including And, Or, Iff, Implies, Ite, and Xor. For each of these operations, we constructed corresponding gates for the AIG. This avoids unnecessary blow-up and retains the semantic structure of the original formula.%

\mparagraph{Circuit to DNF.}
To convert the AIG to DNF in line~\ref{line:allsat}, we use the {\hall}  tool~\cite{FNS23,FNSS24}, which employs a dual-rail based implementation to generate all solutions of the AIG. The solutions are represented as a non-disjoint DNF.

\mparagraph{Constructing the Polytopes.}
While instrumenting {\cvc} to generate the Boolean abstraction, we simultaneously capture which variables correspond to which linear inequalities. For each \textit{cube} of the DNF, therefore, we maintain a mapping indicating which set of linear inequalities this cube corresponds to. Given this map and the cube, we construct the polytope.

\mparagraph{Polytope Volume Computation.}
In {\voled} (line~\ref{line:volume}), we employ the Gaussian cooling-based algorithm developed by \citet{CV16}, which offers a more practical implementation of their theoretically rigorous approach~\cite{CV18}. While the original algorithm~\cite{CV18} provides volume approximation with $(\varepsilon, \delta)$ guarantees, its prohibitive running time of $10^{16}\OO{n^3}$ limits practical applications. The key insight in~\cite{CV16} was that these substantial constant factors can be significantly reduced in practical scenarios without severely compromising accuracy, though this comes at the cost of theoretical guarantees. Our implementation utilizes \textsf{VolEsti}, the software package by Chalkis and Fisikopoulos~\cite{CF21} that implements this practical algorithm.

\mparagraph{Preprocessing}
The polytopes generated by the {\polytope{c}} procedure may contain redundancies and hidden equalities. Hidden equalities render the polytope \textit{degenerate} - resulting in zero volume in $d$ dimensions. Additionally, redundancies significantly complicate the volume computation for the underlying algorithm. We address these challenges by incorporating \textsc{cdd} as a preprocessing step, which effectively identifies hidden equalities and eliminates redundant inequalities. This preprocessing phase yields substantial performance improvements, particularly valuable when processing inequalities derived from SMT files, which often lack optimal formulation.

\mparagraph{Polytope Sampling.}
For polytope sampling algorithm {\GenerateSamples} in line~\ref{line:sample}, we employ the Monte Carlo random walk hit-and-run algorithm~\cite{S84}. Each random walk begins from a point within the convex body and executes a specified number of steps, termed the \textit{walk length}. Greater walk lengths produce final points less correlated with the starting position. The number of steps required to generate an uncorrelated point, one approximately sampled from a distribution, is known as the \textit{mixing time}. Lovász and Vempala~\cite{LV04} showed that $10^{10}\OO{n^2}$ is a  sufficient mixing time for hit and run. However, such requirements are computationally intractable in practice. Following the empirical observations of Lovász and Deák~\cite{LD12}, we therefore constrain our implementation to $n$ steps of hit-and-run, which offers a reasonable balance between sampling quality and computational efficiency.

\mparagraph{Compromises.}
While {\tool} is theoretically sound, the implementation involves a few compromises that prioritize efficiency over strict adherence to theoretical guarantees:

\begin{enumerate}
  \item For volume approximation, algorithms with theoretical guarantees require a running length of $10^{16}\OO{n^3}$ steps, which is impractical. Therefore, in our implementation (line~\ref{line:volume}), we use a practical volume algorithm that employs a convergence-based criterion to determine running length.
  \item The sampling algorithm in line~\ref{line:sample} relies on the hit-and-run method. Known results indicate that achieving independent samples requires $10^{10}n^2$ steps, which is also impractical. However, Lovász and Deák~\cite{LD12} demonstrated that taking $n$ steps of hit-and-run provides practically independent samples. As a result, we use $n$ steps in our implementation.
  \camera{
  Notably, the theoretical guarantees of this sampling algorithm are in $\ell_1$ norm, that is, the samples are guaranteed to be within $\ell_1$ distance of the uniform distribution over the polytope.
  However, {\tool} requires samples to be within $\ell_{\infty}$ distance of the uniform distribution - a stronger requirement than the $\ell_1$ guarantees provided by the hit-and-run sampling algorithm.
  }
\end{enumerate}

\noindent It is worth noting that the state-of-the-art tool, {\shsm} (in the non-exact mode i.e., when relying on {\polyvest}), also makes similar compromises, and therefore, {\ttc} and {\shsm} (in non-exact mode) have similar behavior. The exact mode of {\shsm}, on the other hand, fails to scale to larger instances, as demonstrated in the following section.

\section{Experimental Evaluation}
\label{sec:results}
We evaluated our implementation concerning both  efficiency and  accuracy.

\mparagraph{Baseline.} For performance evaluation, we used the current state of the art volume computation framework {\shsm}\cite{Ge24b}, which offers two distinct modes: the {\polyvest} algorithm, which estimates the volume, and {\vinci}, which performs exact volume computation. To establish meaningful comparisons, we utilized {\polyvest} for performance analysis and {\vinci} for accuracy check. %
\camera{
Another relevant baseline is \citet{ACD20}. Their approach assumes a DNF representation, i.e., a set of polytopes as input; for this purpose we would pass the polytopes generated at line~\ref{line:allsat} of \Cref{alg:ttc}. The publicly available implementation appears to be an early prototype, and in our environment we observed volume discrepancies on some benchmark-derived instances. Resolving these likely requires additional engineering, so we defer a thorough integration and evaluation of this baseline to future work.
  }

\mparagraph{Benchmarks.}
As a first step, we sought to rely on benchmarks from SMT-Lib, but these benchmarks could not be handled by {\polyvest} or {\vinci} owing to them containing extremely thin geometric regions where dimensions may be constrained to narrow ranges (e.g., (0,10$^{-7}$)). In these cases, {\polyvest} and {\vinci} fail to handle the required precision and incorrectly classify these polytopes as degenerate with zero volume.
To avoid results confounded by numerical precision rather than algorithmic differences, we do not include  SMT-LIB in our evaluation.
Accordingly, we focus on the construction of synthetic instances that are  disjunctions of intersecting polytopes, with each polytope defined in H-representation ($Ax \leq b$). In total, our benchmark suite consists  of {\tot} benchmarks.

\begin{enumerate}
  \item We vary two parameters: (1) dimension parameter $n$: ranges from 6 to 34, (2) number of polytopes $m$: ranges from 6 to 42.
  \item For each instance, we generate polytopes using  three different geometric shapes studied by \citet{CV16}:
 (a) \textit{Cubes:} $n$-dimensional cubes with random bounds, then translated and rotated. (b) \textit{Zonotypes: } Minkowski sum of $n$ different $d$-dimensional vectors generated randomly. (c) \textit{Simplex: } Shapes where all coordinates are nonnegative and sum to at most 1, defined as $\{x \in \mathbb{R}^n : \sum_{i=1}^n x_i \leq 1, x_i \geq 0\}$.
\end{enumerate}

\mparagraph{Environment.} We conducted all our experiments on a high-performance computer cluster, with each node consisting of Intel Xeon Gold 6148 CPUs. We allocated one CPU core and a 5GB  memory limit to each solver instance pair. To adhere to the standard timeout used in model counting competitions, we set the timeout for all experiments to 3600 seconds. We use values of $\varepsilon=0.8$ and $\delta=0.2$, in line with prior work in the model counting community.

\vspace{.5cm}
\noindent
With the above setup, we conduct extensive experiments to understand the following:
\begin{enumerate}[font=\bfseries RQ, leftmargin=\widthof{[RQ]}+\labelsep]
  \item How does the runtime performance of {\tool} compare to that of {\base}?
  \item How does the performance of {\tool} scale with different benchmark parameters?
  \item How accurate is the count computed by {\tool} in comparison to the exact count?

\end{enumerate}

\begin{table}[tbh]
  \centering
  \begin{tabular}{lcc}
    \hline
 Solver & Solved & PAR-2 \\
    \hline
 {\vinci} & {\solvdvinci} & 6097.63 \\
 {\polyvest} & {\solvdbase} & 6199.73 \\
 {\ttc} & \textbf{\solvdttc} & \textbf{255.48} \\
    \hline
    \end{tabular}
  \caption{Performance comparison on {\tot} instances.}
  \label{tab:performance}
\end{table}

\mparagraph{Summary of Results.}  {\tool} achieves a significant performance improvement over {\base} by finishing on {\solvdttc} instances in a benchmark set consisting of {\tot}, while {\base} could only finish on {\solvdbase} instances. {\base} barely finishes on instances with more than 25 polytopes or 20 dimensions, while {\ttc} seamlessly handles  40 polytopes of 35 dimensions. The accuracy of the approximate count is also noteworthy, with an average error of a count by {\tool} of only 0.{\accperc}\footnote{All benchmarks and logfiles are available at \url{https://doi.org/10.5281/zenodo.16782810}}. %

\subsection{Performance of {\tool}}

\mparagraph{Instances Solved.} In \Cref{tab:performance}, we compare the number of benchmarks that can be solved by {\base} and {\tool}. First, it is evident that the {\base} only solved {\solvdbase} out of the {\tot} benchmarks in the test suite, indicating its lack of scalability. Conversely, {\tool} solved {\solvdttc} instances, demonstrating a substantial improvement compared to {\base}.

\begin{figure}[!htb]
  \centering
  \begin{subfigure}[b]{0.6\linewidth}
    \centering
    \resizebox{\linewidth}{!}{\input{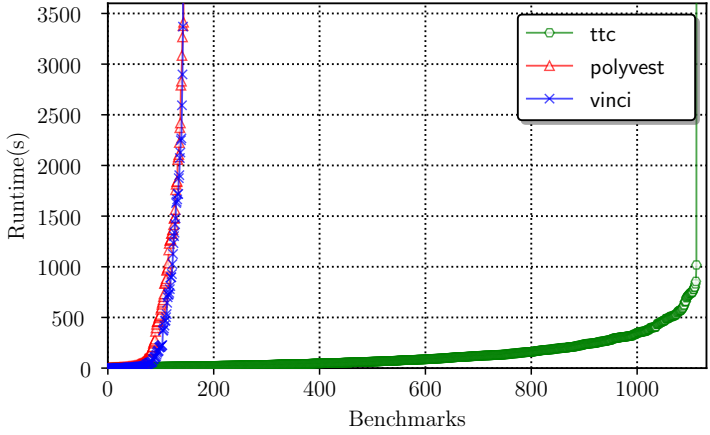}}
    \caption{Cactus plot comparing runtime of different tools.}
    \label{fig:cactus}
  \end{subfigure}
  \hfill
  \begin{subfigure}[b]{0.36\linewidth}
    \centering
    \resizebox{\linewidth}{!}{\input{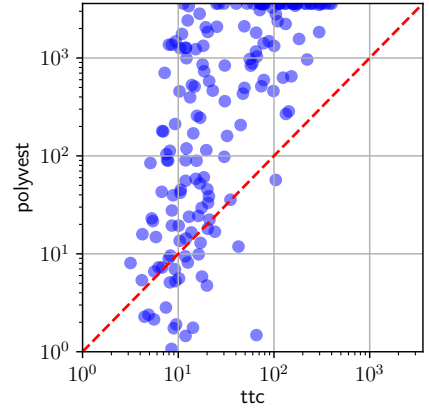}}
    \caption{Runtime comparison of {\ttc} w.r.t. {\polyvest}.}
    \label{fig:scatter}
  \end{subfigure}
  \caption{Performance comparison: (a) cactus plot of runtimes, (b) scatter plot comparing {\ttc} and {\polyvest}.}
  \label{fig:perf-combined}
\end{figure}

\mparagraph{Solving Time Comparison.}  A performance evaluation of {\base} and {\tool} is depicted in \Cref{fig:cactus}, which is a cactus plot comparing the solving time. The $x$-axis represents the number of instances, while the $y$-axis shows the time taken. A point $(i, j)$ in the plot represents that a solver solved $j$ benchmarks out of the {\tot} benchmarks in the test suite in less than or equal to $j$ seconds. The curves for {\base} and {\tool} indicate that for a few instances, {\base} was able to give a quick answer, while in the long run, {\tool} could solve many more instances given any fixed timeout.

In \Cref{tab:performance} we also show the PAR-2 score of the solvers, which is the mean runtime over all instances, assigning a cost of $2T$ to each instance timed out at $T$. {\ttc} shows significantly small PAR-2 score.
In \Cref{fig:scatter}, we present a comparative analysis of solving times between {\ttc} and {\base}. Each data point $(x,y)$ represents an instance that was solved in $x$ seconds by {\ttc} and $y$ seconds by {\base}. Points appearing below the dotted red diagonal line indicate instances where {\base} outperformed {\ttc} in solving time. {\ttc} demonstrates superior performance on the vast majority of instances.%

In \Cref{tab:performance} we also show the PAR-2 score of the solvers, which is the mean runtime over all instances, assigning a cost of $2T$ to each instance timed out at $T$. {\ttc} shows significantly small PAR-2 score.
In \Cref{fig:scatter}, we present a comparative analysis of solving times between {\ttc} and {\base}. Each data point $(x,y)$ represents an instance that was solved in $x$ seconds by {\ttc} and $y$ seconds by {\base}. Points appearing below the dotted red diagonal line indicate instances where {\base} outperformed {\ttc} in solving time. {\ttc} demonstrates superior performance on the vast majority of instances.%

\camera{
\mparagraph{Time utilization in different components.} Theoretically, AIG to DNF conversion can take exponential amount of time. However, in our experiments, the DNF conversion time was negligible. The maximum time spent on this conversion was 1.2 seconds, with most instances completing the conversion in less than one second. The majority of the computation time was spent on calculating the individual volumes of the polytopes. For all instances that took more than 100 seconds to solve, over 60\% of the time was dedicated to polytope volume computation.
}

\mparagraph{Experiment with different $\varepsilon$.}
\Cref{table:diffeps} reports the runtime of {\ttc} for varying values of $\varepsilon$. The runtime remains largely consistent across different settings. Notably, with $\varepsilon = 0.1$, {\ttc} solves only two fewer instances compared to $\varepsilon = 0.8$, indicating that {\ttc} maintains its efficiency even at lower values of $\varepsilon$.

\begin{figure}[!tbp]
  \centering
  \begin{subfigure}[b]{0.48\linewidth}
    \centering
    \resizebox{\linewidth}{!}{\input{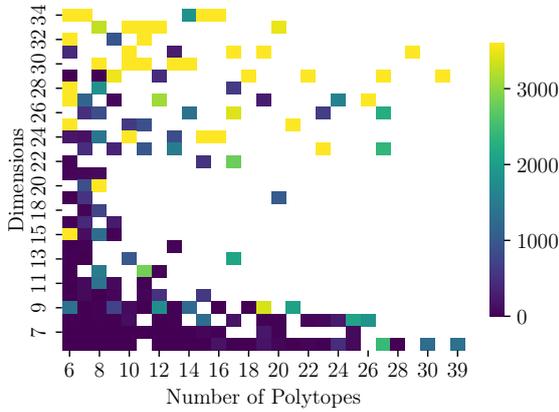}}
    \caption{{\polyvest}}
    \label{fig:scaling-shsm}
  \end{subfigure}
  \hfill
  \begin{subfigure}[b]{0.48\linewidth}
    \centering
    \resizebox{\linewidth}{!}{\input{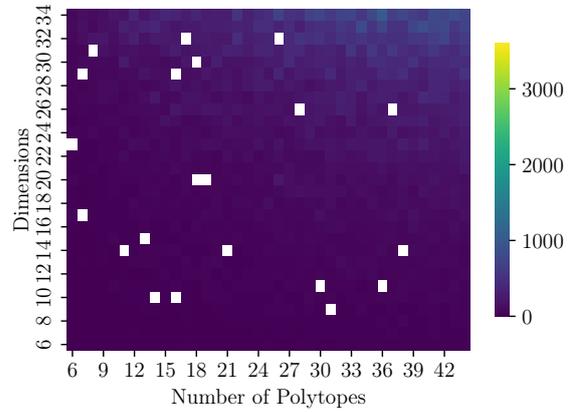}}
    \caption{{\ttc}}
    \label{fig:scaling-ttc}
  \end{subfigure}
  \caption{Time taken w.r.t. dimensions and \# polytopes.}
  \label{fig:scaling-comparison}
\end{figure}

\subsection{Scaling}
In \Cref{fig:scaling-shsm} and \ref{fig:scaling-ttc}, we evaluate the scalability of {\ttc} and {\base} with respect to both the number of polytopes and dimensions. The plots are organized with the number of polytopes increasing from left to right along the $x$-axis, while the number of dimensions increases from top to bottom along the $y$-axis. Each pixel corresponds to a specific instance in our benchmark dataset, with the color intensity representing the solver's runtime performance.
As demonstrated in \Cref{fig:scaling-shsm}, {\base} exhibits significant performance degradation when handling instances exceeding 12 polytopes or 12 dimensions. By contrast, \Cref{fig:scaling-ttc} reveals that {\ttc} efficiently processes configurations with up to 40 polytopes and 35 dimensions without notable performance deterioration. %

\subsection{Quality of Approximation}

\begin{figure}[!tbp]
  \centering
  \resizebox{0.6\linewidth}{!}{\input{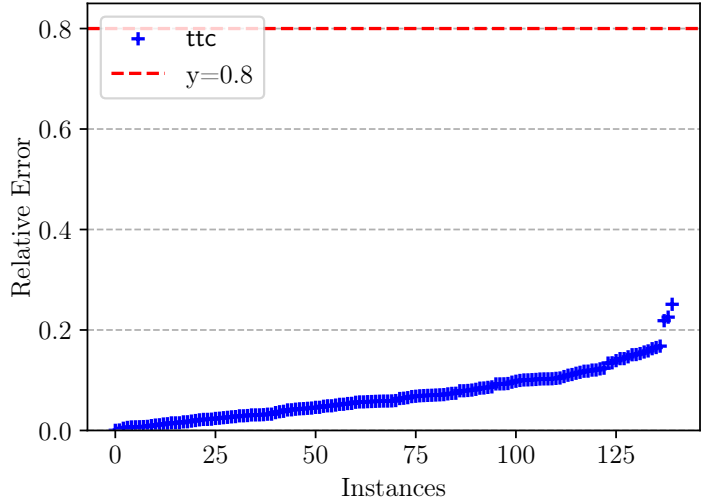}}
  \caption{Quality of approximation: observed error.}
  \label{fig:error-ttc-vinci}
\end{figure}

In our experimental evaluation, we found the exact volume of {\solvdvinci} benchmarks from {\vinci}, enabling us to calculate the error made by {\ttc} on these instances.
We quantify the  error made  by {\tool} by the parameter $e = \frac{|b - s|}{b}$, where $b$ represents the count from {\vinci} and $s$ from {\tool}. This measure is the \textit{observed error}, analogous to the theoretical error guarantees provided by {\ttc}. Analysis of all {\solvdvinci} cases found the median $e$ to be $0.059$, geometric mean $0.038$, and maximum $0.39$, contrasting sharply with a theoretical guarantee of $0.8$. This signifies {\tool} substantially outperforms its theoretical bounds.
In \Cref{fig:error-ttc-vinci} we plot the observed error, where $x$-axis, we have the benchmarks, and on the $y$-axis we have the observed errors. The observed error is below $0.2$ for most of the instances.

In \Cref{tab:error-obs-vs-theo} we showed different observed errors when we run {\ttc} with different $\varepsilon$ values. The median and maximum observed errors decrease with the theoretical $\varepsilon$. The maximum \textit{observed error} with  $\varepsilon = 0.1$, is greater than theoretical, which is not unnatural, given the $(\varepsilon, \delta)$ guarantee nature.

\begin{table}[h]
\centering
\begin{tabular}{llccc}
  \hline
  Solver & Solved & PAR-2 \\
  \hline
  {\vinci} & {\solvdvinci} & 6097.63 \\
  {\polyvest} & {\solvdbase} & 6199.73 \\
  {\ttc}-0.1 & 1110 & 275.52 \\
  {\ttc}-0.4 & 1113 & 248.04 \\
  {\ttc}-0.8 & {\solvdttc} & 255.48 \\
  \hline
  \end{tabular}
  \caption{\centering Instances solved with different $\varepsilon$ for {\ttc}.}
  \label{table:diffeps}
\end{table}

\begin{table}[!tbp]
  \centering
  \begin{tabular}{@{}lllll@{}}
    \toprule
    \multicolumn{2}{l}{Theoretical}           & 0.1  & 0.4 & 0.8 \\ \midrule
    \multirow{2}{*}{Observed} & Median    & 0.03    & 0.04   & 0.04  \\
    & Max     & 0.22  & 0.20   & 0.39 \\\bottomrule
  \end{tabular}
  \caption{Theoretical vs. observed error at different $\varepsilon$.}
  \label{tab:error-obs-vs-theo}
\end{table}

\label{subsec:performance}

\section{Conclusion}
\label{sec:concl}
This paper introduces {\ttc}, a scalable approximate SMT volume computation tool that demonstrates exceptional performance on practical benchmarks. Our approach harnesses probabilistic techniques to deliver theoretical guarantees on computation results,
and empirical results significantly surpassing theoretical guarantees.
Our work suggests several promising research directions. First, many formulas contain equality constraints that result in zero volume when computing in $d$ dimensions. A natural extension would be to develop methods for correctly computing $(d-k)$-dimensional volume in such cases. Second, while we prioritized performance over strict theoretical guarantees in our implementation, experimental results consistently demonstrate error rates well below theoretical bounds. This raises the intriguing question, whether rigorous guarantees can be established for our current implementation without sacrificing its performance advantages.

\section*{Acknowledgement}
We are thankful to Sourav Chakraborty, Radoslav Dimitrov, Arijit Ghosh and Mate Soos for the many useful discussions.
This work was supported in part by the Natural Sciences and Engineering Research Council of Canada (NSERC) [RGPIN-2024-05956]. The work was done when Arijit Shaw was a visiting graduate student at the University of Toronto. Uddalok Sarkar was supported by Google PhD fellowship. Computations were performed on the Niagara supercomputer at the SciNet HPC Consortium. SciNet is funded by Innovation, Science and Economic Development Canada; the Digital Research Alliance of Canada; the Ontario Research Fund: Research Excellence; and the University of Toronto.

\bibliographystyle{alpha}
\bibliography{bib}
\clearpage
\appendix
\onecolumn
\section{Additional Proofs}

\process*
\begin{proof}
  Equivalence of (1) and (2) follows from Claim 2 of \cite{soos2024engineering}.
  Therefore we restrict ourselves to proving the equivalence of (1) and (3).
  Let $\zeta$ denote the multiplicative deviation in the probability of selecting $s_i$ in a single draw from the $\eta$-close uniform distribution, i.e., $\zeta \in [1-\eta, 1+\eta]$. Consider the process in (1): we draw $N \sim \pois(\sz \cdot p/2)$ samples with replacement from $S$, each according to the $\eta$-close uniform distribution. The probability that $s$ is chosen exactly $r$ times is:
  \begin{align*}
    &\Pr\left(\text{Choosing } s \text{ exactly $r$ many times}\right) \\
    =& \sum_{k=r}^{\infty} {k\choose{r}} \left(\frac{\zeta}{|S|}\right)^r \left(1 - \frac{\zeta}{|S|}\right)^{k-r} \cdot \Pr(N = k)\\
    =& \sum_{k=r}^{\infty} {k\choose{r}} \left(\frac{\zeta}{|S|}\right)^r \left(1 - \frac{\zeta}{|S|}\right)^{k-r} \cdot \frac{(\frac{p}{2}\sz)^k e^{-\frac{p}{2}\sz}}{k!} && \text{Since } N \sim \pois(\sz \cdot p/2)\\
    =& \sum_{k=r}^{\infty} {k\choose{r}} \left(\frac{\zeta}{|S|}\right)^r \left(1 - \frac{\zeta}{|S|}\right)^{k-r} \cdot \frac{(\frac{p'}{2}|S|)^k e^{-\frac{p'}{2}|S|}}{k!} && \text{Let } p' = \sz \cdot p / |S|\\
    =& \frac{(\zeta\cdot p'/2)^r e^{-\frac{p'}{2}|S|}}{r!} \sum_{k=r}^{\infty} \frac{(p'/2)^{k-r}}{(k-r)!} \left(|S| - \zeta\right)^{k-r}\\
    =& \frac{(\zeta\cdot p'/2)^r e^{-\frac{p'}{2}|S|}}{r!} \sum_{k=0}^{\infty} \frac{(p'/2)^{k}}{k!} \left(|S| - \zeta\right)^{k}\\
    =& \frac{(\zeta\cdot p'/2)^r e^{-\frac{p'}{2}|S|}}{r!} e^{\frac{p'}{2}(|S| - \zeta)}\\
    =& \frac{(\zeta\cdot p'/2)^r e^{-\frac{p'}{2}|S| + \frac{p'}{2}|S| - \zeta\cdot\frac{p'}{2}}}{r!}\\
    =& \frac{(\zeta\cdot p'/2)^r e^{-\zeta\cdot\frac{p'}{2}}}{r!} 
  \end{align*}
  Observe that, $\alpha p \leq p' \leq \beta p$. 
  Therefore the probability of choosing $s$ exactly $r$ many times is $\frac{(\zeta\cdot p'/2)^t e^{-\zeta\cdot \frac{p'}{2}}}{r!}$, which is the same as the probability of drawing $r \sim \pois(\apxp/2)$ and adding $r$ copies of $s_i$ to $\cX$ where $\apxp \in [(1-\eta)\alpha p, (1+\eta)\beta p]$. Now we move towards proving the independence. To do that we consider the set $S$ to be $S = \{s_1, \ldots, s_l\}$. We assume that the probability of choosing $s_i$ in a single draw is $\frac{\zeta_i}{|S|} \in \left[\frac{1-\eta}{|S|}, \frac{1+\eta}{|S|}\right]$ for each $i \in [l]$. Hence, note that $\sum_i \zeta_i = |S|$. Consequently,
  \begin{align*}
    &\Pr\left(\bigwedge_{i=1}^l\text{Choosing } s_i \text{ exactly $r_i$ many times}\right) \\
    =& \frac{k!}{r_1! \ldots r_l!} \left(\frac{\zeta_1}{|S|}\right)^{r_1} \ldots \left(\frac{\zeta_l}{|S|}\right)^{r_l} \cdot \Pr(N = k) && \text{Where } k = r_1 + \ldots + r_l\\
    =& \frac{k!}{r_1! \ldots r_l!} \left(\frac{\zeta_1}{|S|}\right)^{r_1} \ldots \left(\frac{\zeta_l}{|S|}\right)^{r_l} \cdot \frac{(\frac{p}{2}\sz)^k e^{-\frac{p}{2}\sz}}{k!}\\
    =& \frac{1}{r_1! \ldots r_l!} \left(\frac{\zeta_1}{|S|}\right)^{r_1} \ldots \left(\frac{\zeta_l}{|S|}\right)^{r_l} \cdot (\frac{p'}{2}|S|)^k e^{-\frac{p'}{2}|S|} && \text{Let } p' = \sz \cdot p / |S|\\
    =& \frac{1}{r_1! \ldots r_l!} \left(\zeta_1\cdot p'/2\right)^{r_1} \ldots \left(\zeta_l\cdot p'/2\right)^{r_l} e^{-\frac{p'}{2}|S|} \\
    =& \frac{1}{r_1! \ldots r_l!} \left(\zeta_1\cdot p'/2\right)^{r_1} \ldots \left(\zeta_l\cdot p'/2\right)^{r_l} e^{-\frac{p'}{2}(\zeta_1 + \ldots + \zeta_l)} && \text{Since } \zeta_1 + \ldots + \zeta_l = |S|
  \end{align*}
  The above expression is exactly equal to the probability of independently drawing $r_i \sim \pois(\apxp/2)$, that is, drawing $r_i$ with probability $\frac{(\zeta_i\cdot p'/2)^{r_i}e^{-\zeta_i\cdot \frac{p'}{2}}}{r_i!} $ for each $i \in [l]$, and adding $r_i$ copies of $s_i$ to $\cX$. 
\end{proof}

\helper*

\begin{proof}
  \begin{align*}
    k^n &= \left(\frac{\tfrac{16n}{\eta} \sqrt{\log \tfrac{4r}{\eta}} + \kappa}{\tfrac{16n}{\eta} \sqrt{\log \tfrac{4r}{\eta}} -  \kappa}\right)^n = \left(\frac{\tfrac{16n}{\eta} \sqrt{\log \tfrac{4r}{\eta}} + \sqrt{2 \log \frac{4}{\eta}}+\sqrt{2\log r}
    }{\tfrac{16n}{\eta} \sqrt{\log \tfrac{4r}{\eta}} -  \sqrt{2 \log \frac{4}{\eta}}-\sqrt{2\log r}
    }\right)^n
  \end{align*}
  Now let us focus on the term $\sqrt{2 \log \frac{4}{\eta}}+\sqrt{2\log r}$. We can bound this term as follows:
  \begin{align*}
    \sqrt{2 \log \frac{4}{\eta}}+\sqrt{2\log r} &\leq 2\sqrt{\frac{\log \frac{4}{\eta} + \log r}{2}} && \text{Since $\sqrt{x} + \sqrt{y} \leq 2\sqrt{\frac{x+y}{2}}$}\\
    &= 2\sqrt{\log \frac{4r}{\eta}} && \text{Since $\log a + \log b = \log(ab)$}
  \end{align*}
  Therefore, we can upper bound the term $k^n$ as follows:
  \allowdisplaybreaks
  \begin{align*}
    k^n &\leq \left(\frac{\tfrac{16n}{\eta} \sqrt{\log \tfrac{4r}{\eta}} + 2\sqrt{\log \frac{4r}{\eta}}}{\tfrac{16n}{\eta} \sqrt{\log \tfrac{4r}{\eta}} - 2\sqrt{\log \frac{4r}{\eta}}}\right)^n\\
    &= \left(\frac{n + \tfrac{\eta}{8}}{n - \tfrac{\eta}{8}}\right)^n \\
    &= \left(1 + \frac{\tfrac{\eta}{4}}{n - \tfrac{\eta}{8}}\right)^n \\
    &\leq \exp\left(\frac{n \cdot \tfrac{\eta}{4}}{n - \tfrac{\eta}{8}}\right) && \text{Since $1+x \leq e^x$ for $x\geq 0$}\\
    &= \exp\left(\tfrac{\eta}{4} \cdot \left(1 + \frac{\tfrac{\eta}{8}}{n -\tfrac{\eta}{8}}\right)\right)\\
    &\leq \exp\left(\tfrac{\eta}{2}\right) && \text{Since for $\eta < n$ we have $\frac{\tfrac{\eta}{8}}{n -\tfrac{\eta}{8}} \leq 1$}\\
    &\leq \frac{1+\eta/4}{1-\eta/4} && \text{For $\eta < 1$}
  \end{align*}
  The last inequality follows from the fact that for $0 < x < 1$, we have $\exp(2x) \leq \frac{1+x}{1-x}$. This completes the proof.
\end{proof}

\end{document}